# Effect of Edge Roughness in Graphene Nanoribbon Transistors

Youngki Yoon and Jing Guo*

Department of Electrical and Computer Engineering,

University of Florida, Gainesville, Florida, 32611

## ABSTRACT

The effects of edge irregularity and mixed edge shapes on the characteristics of graphene nanoribbon transistors are examined by self-consistent atomistic simulations based on the non-equilibrium Green's function formalism. The minimal leakage current increases due to the localized states induced in the band gap, and the on-current decreases due to smaller quantum transmission and the self-consistent electrostatic effect in general. Although the ratio between the on-current and minimal leakage current decreases, the transistor still switches even in the presence of edge roughness. The variation between devices, however, can be large, especially for a short channel length.

*e-mail: guoj@ufl.edu

Graphene is a two-dimensional (2D) atomically-thin layer of carbon atoms, which has been a subject of strong interest for potential nanoelectronics applications recently due to its excellent carrier transport properties.[1] By using a state-of-the-art patterning technique like e-beam lithography, a narrow strip of one-dimensional graphene nanoribbon (GNR)[2,3] is possible from a 2D graphene monolayer.[4,5] Theoretical studies on the GNR field-effect transistors (FETs) with perfect edges have been reported.[6-9] The understanding of edge effects on the device performance, however, is essential because the state-of-the-art lithography technique is far from the atomistic precision. The important role of edge roughness on the electronic structures of GNRs has been examined on the material level[10,11], but the effects on the device performance of GNRFETs remain mostly unknown. In this letter, the effects of edge roughness on the device characteristics of GNRFETs are studied by self-consistent atomistic simulations. The results indicate that the carrier transport and self-consistent electrostatics of GNRFETs are strongly related to the GNR edge shapes, which significantly affect device physics and transistor characteristics.

Two types of imperfect edges are induced and examined based on a bottom-gated GNR Schottky barrier (SB) FET with an armchair-edge GNR (A-GNR) channel, as shown in Fig. 1(a), at room temperature ($T$=300 K). The $N$=12 A-GNR channel has a band gap of $E_g \approx 0.83$ eV, where $N$ is GNR index that denotes the number of carbon dimer lines of A-GNR.[3] Because the GNR edges are typically a mixture of armchair and zigzag edge shapes, a zigzag-edge GNR (Z-GNR) portion is inserted in the middle of the A-GNR channel as the first type of imperfect edges. Fig. 1(a) also shows the middle part of this mixed-edge GNR (M-GNR) channel. The second type of imperfect edges takes edge irregularity into account by randomly adding or removing atoms from the edges of the A-GNR channel as described in detail later. The following device parameters are used in the simulation. A 20-nm-long GNR channel is connected to metal source and drain[4,5], and Schottky barriers form between the channel and the source (drain) contacts. The SB height between the metal source (drain) and the GNR channel is one half of the band gap, $\Phi_{Bn} = \Phi_{Bp} = E_g/2$, and flat band voltage is zero. A SiO$_2$ ($\kappa \approx 4$) gate insulator with a thickness of $t_{ox}$=2 nm is used. A power supply voltage of $V_{DD}$=0.5 V is used and source electrode is grounded, which results in zero source voltage.

The DC characteristics of ballistic GNR SBFETs are simulated by solving the Schrödinger equation using the non-equilibrium Green's function (NEGF) formalism[12] self-consistently with a three-dimensional (3D) Poisson equation. A tight binding Hamiltonian with a p$_z$ orbital basis set is used for GNR channel with imperfect edges, and a p$_z$ orbital coupling parameter of $t_0$=3 eV is assumed. The edge bond relaxation and spin-polarized zigzag edges predicted by *ab-initio* calculations[13] add perturbations to the band structure. The effects, however, are expected to be small under high bias conditions and are not expected to change the qualitative conclusions. We, therefore, neglect these effects for efficient device simulations. A phenomenological description of the metal contacts, which assumes a continuous and constant contact density of states in the energy range of interest, is used to compute the source and drain self-energies.[14] A 3D Poisson equation is numerically solved using the finite element method (FEM) to treat self-consistent electrostatics.

Fig. 1(b) plots $I_D$ vs. $V_G$ characteristic at $V_D=V_{DD}=0.5$ V with a M-GNR channel (solid lines), compared to that of a $N=12$ A-GNRFET (dashed lines). The minimal leakage current, which is defined as the minimum in the $I_D$ vs. $V_G$ curve, is achieved at $V_G=V_{min}=V_D/2$ because of the ambipolar behavior of SBFETs. The off-state, therefore, is chosen at $V_G=V_{min}=V_D/2$, which could be shifted to zero gate voltage by properly choosing the gate metal material and engineering the flat band voltage. The on-current $I_{on}$ is defined at $V_G=V_{min}+V_{DD}$.

Fig. 1(b) shows that the minimal leakage current $I_{min}$ of M-GNRFET with $L_z=0.43$ nm is a factor of 3.5 larger than that of a perfect edge GNRFET. Fig. 1(c) indicates that even a short Z-GNR portion induces localized states in the band gap energy range, which facilitate quantum-mechanical tunneling between the source and the drain. At the minimal leakage bias condition, the quantum-tunneling current through the localized states in the band gap energy is larger than the current carried by the conduction and the valence bands. The increase of the minimal leakage current, therefore, is significant so that on-off ratio is reduced from 2600 to 550.

On the other hand, the on-current of a M-GNRFET is 30% smaller than that of a $N=12$ A-

GNRFET. Fig. 1(d) shows the LDOS of the M-GNRFET at on-state with self-consistent potential profile (dashed lines). The on-current of the M-GNRFET decreases due to (i) a smaller quantum transmission through the imperfect channel and (ii) change of the self-consistent potential. In order to separate these two effects, we also computed the current of the M-GNRFET non-self-consistently using the same potential profile as the perfect edge GNRFET. Compared to the on-current of the perfect edge GNRFET, 1.91 µA, the smaller quantum transmission reduces the on-current to $I_{nsc}$= 1.05 µA, and the self-consistent electrostatic effect alters it to 1.36 µA. It is clear that the smaller quantum transmission is mainly responsible for the decrease of the on-current. In order to examine the effect of a longer zigzag portion, we also simulated the case with a longer $L_z$=2.13 nm, which has an on-off ratio of ~500 and still shows a considerable current modulation by the applied gate voltage.

Next, the effects of the edge irregularity of GNR on the performance of SBFETs are studied. Fig. 2(a) shows an example of the irregular-edge GNR (I-GNR). The edge of I-GNR is generated by the following way. (i) The probability of edge irregularity is determined by the input parameter $0 \leq P \leq 1$. (ii) Starting from a perfect $N$=12 A-GNR, the positions to add or remove a pair of carbon atoms on the edge are randomly selected along the transport direction. (iii) If it is placed at the convex edge position of the perfect A-GNR, two carbon atoms on that edge are removed, and if at the concave edge position, two C atoms are added. (iv) The modification of the perfect A-GNR edge at the consecutive positions is avoided to generate a stable edge shape.

Fig. 2(b) and 2(c) show the LDOS's of two I-GNRFETs, which are generated by specifying the same $P$=0.06 in two computer simulations under zero bias. We refer to the I-GNR of Fig. 2(b) and Fig. 2(c) as case 1 and case 2, respectively, in the subsequent discussions. It is noticed that a large LDOS in the band gap energy range is induced by the edge irregularity for case 1 throughout the whole channel. Even with the same probability of the channel irregularity, the variation of the LDOS between case 1 and case 2 is large, which should lead to a large variation between the devices. To examine the effect of a larger probability of edge irregularity, Fig. 2(c) plots the LDOS of another I-GNR structure generated with $P$=0.12 (referred to as case 3).

In order to explore the effect on device characteristics, we simulated the $I_D$-$V_G$ characteristics of the above three cases, as shown in Fig. 3(a) and 3(b). The minimal leakage current of case 1 is one order larger than that of the perfect A-GNRFET due to the quantum-mechanical tunneling through the states induced in the band gap energy range, as shown in Fig. 3(c) and Fig. 4(a). Note that the increase of $I_{min}$ by the gap states is severe only for a relatively short channel. If the channel length is much longer than characteristic localization length of the gap states, I-GNR becomes an Anderson insulator and the gap states hardly facilitate quantum tunneling through the band gap region.[10] For the same channel length of GNRs, the leakage current of the GNR with a larger width is more sensitive to edge shape because it has a larger characteristic localization length.[10] On the other hand, the on-current of case 1 is about 90 % of the perfect edge case, which is caused by both the self-consistent electrostatic and the quantum transport effects. The bands of the case 1 flow up because the gap states induced by the edge irregularity increase the electron density at on-state throughout the channel. The flow-up of the bands leads to a smaller energy range for electrons to tunnel from the source to the channel, as shown in Fig. 4(b), which tends to decrease the current. At the same time, the band gap states at the beginning of the channel facilitate quantum tunneling through the Schottky barrier at the source end, which leads to an increase of tunneling probability as shown in Fig. 4(b) and tends to increase the current. These two effects compensate each other to a certain extent and the on-current of the case 1 I-GNRFET is about 90% of that of the perfect edge GNRFET.

It is noticed that the device performance can have a large variation between two I-GNRFETs generated by specifying the same probability of edge irregularity. The minimal leakage current of case 2 is one order smaller than that of case 1. The on-current of case 2 is also much smaller than that of the perfect A-GNRFET or case 1, which is caused mostly by the degradation of transport property as for M-GNRFETs. Two examples with $P=0.06$ in this study show that the variation of device performance may be large even under the same level of edge precision control, especially when the channel length is short. If the probability is further increased to $P=0.12$ (case 3), the on-off ratio is reduced to ~40, but it still gives a transistor switching behavior.

In conclusion, we have simulated the effects of mixed and irregular edge shapes on the device performance of GNR SBFETs. Even a short length of mixed edge reduces the on-off ratio, with increased $I_{min}$ by the localized gap states and reduced $I_{on}$ due to the smaller transmission probability. In the presence of edge irregularity, GNR SBFETs show a large variation in terms of the device performance. Although the $I_{on}/I_{min}$ is degraded, the gate still modulates the source-drain current considerably even in the presence of edge roughness.

**FIGURE CAPTIONS**

Fig.1 (a) The simulated device of a bottom gate Schottky barrier (SB) field-effect transistor (FET), and the middle of a mixed-edge graphene nanoribbon (M-GNR) structure as a channel material. The length of zigzag edge portion is $L_z \approx 0.43$ nm, which is located at the middle of the GNR channel in the longitudinal direction. (b) $I_D$-$V_G$ characteristics for a perfect armchair-edge GNR (A-GNR) FET (dashed line) and a M-GNRFET (solid line) in a log scale and a linear scale. (c) *Local density-of-states (LDOS):* Energy-resolved density of states (DOS) versus channel position at off-state ($V_G$=$V_{min}$=0.25 V, $V_D$=0.5 V), and (d) at on-state ($V_G$=$V_{min}$+$V_{DD}$=0.75 V, $V_D$=0.5 V) for the M-GNRFET. In (c) and (d), $E_C$ and $E_V$ of a $N$=12 A-GNRFET are plotted with the solid lines, and the dashed lines are plotted by $E_m(x) \pm E_g/2$, where $E_m(x)$ is self-consistent electrostatic potential with M-GNR along the channel position and $E_g$ is the band gap of the $N$=12 A-GNR.

Fig. 2 (a) One example of the atomistic configuration of an irregular-edge GNR (I-GNR) with a probability of edge irregularity, $P$=0.12. (b) LDOS of an I-GNR with $P$=0.06 (case 1). (c) LDOS of another I-GNR with $P$=0.06 (case 2). Because a random process generates the edges of simulated I-GNRs, different I-GNRs with different edge shapes are possible even with the same $P$. (d) LDOS for an I-GNR with $P$=0.12 (case 3). Solid lines are $E_C$ and $E_V$ of the $N$=12 A-GNR.

Fig. 3 $I_D$-$V_G$ characteristics for a perfect edge GNRFET and three I-GNRFETs (a) in a log scale and (b) in a linear scale. (c) LDOS at off-state ($V_G$=$V_{min}$≈0.3 V, $V_D$=0.5 V) and (d) at on-state ($V_G$=$V_{min}$+$V_{DD}$≈0.8 V, $V_D$=0.5 V) for the case 1 I-GNRFET. In (c) and (d), the solid lines and the dashed lines are same as explained in the Fig. 1 caption.

Fig. 4 (a) Energy-resolved current spectrum, $J_E(E)$, at off-state ($V_G \approx 0.3$ V, $V_D=0.5$ V) and (b) at on-state ($V_G \approx 0.8$ V, $V_D=0.5$ V) for the case 1 I-GNRFET. The solid lines are for the case 1 I-GNRFET, and the dashed lines are for the perfect $N=12$ A-GNRFET.

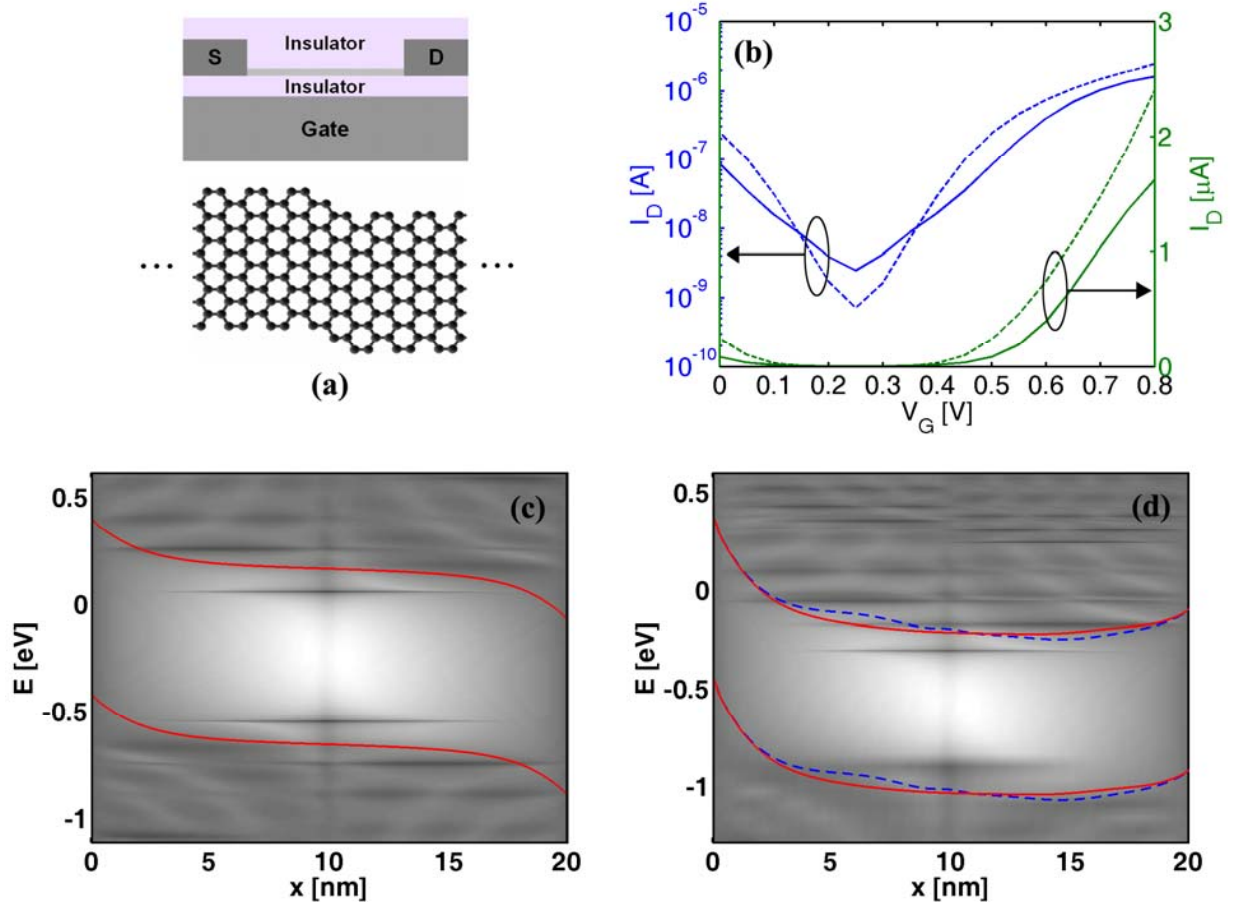

Fig. 1

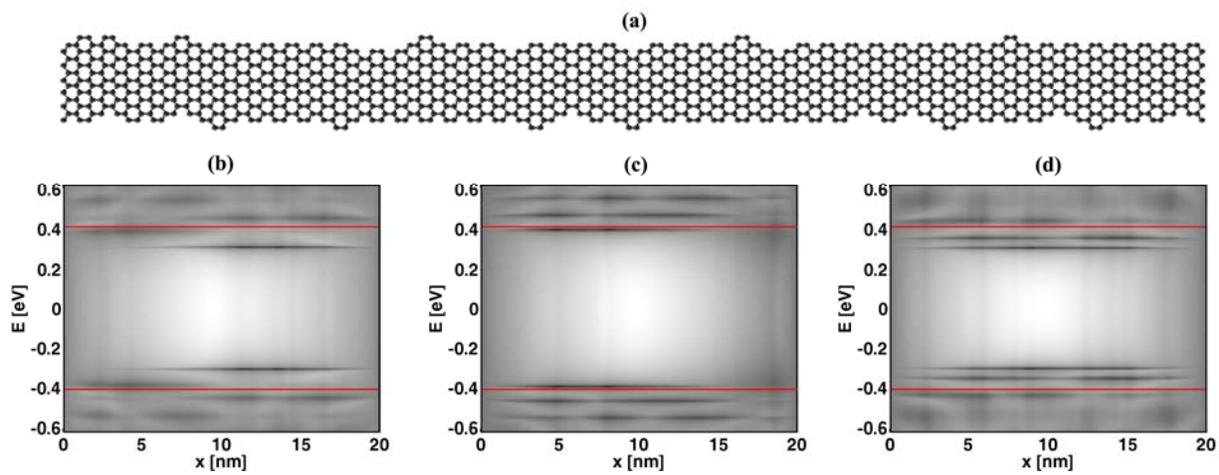

Fig. 2

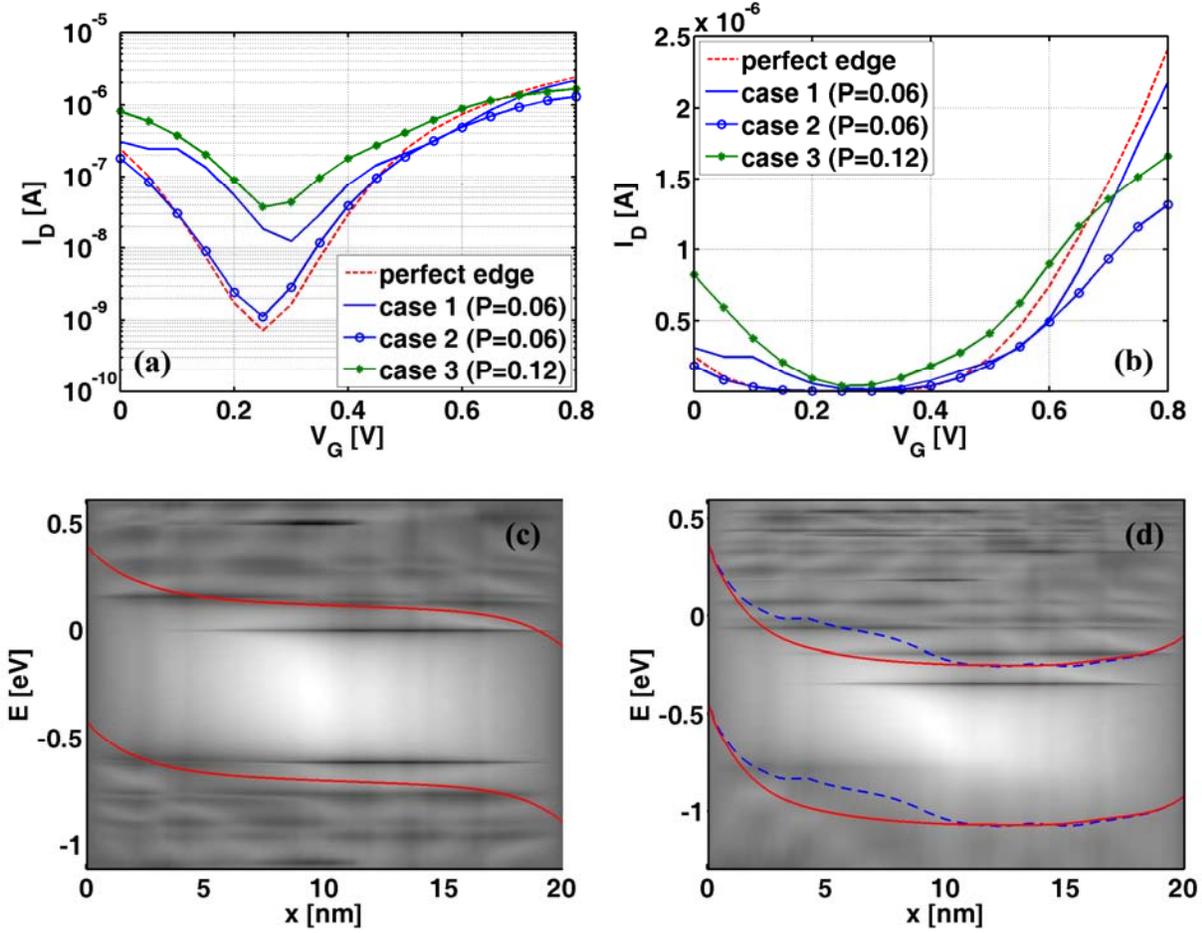

Fig. 3

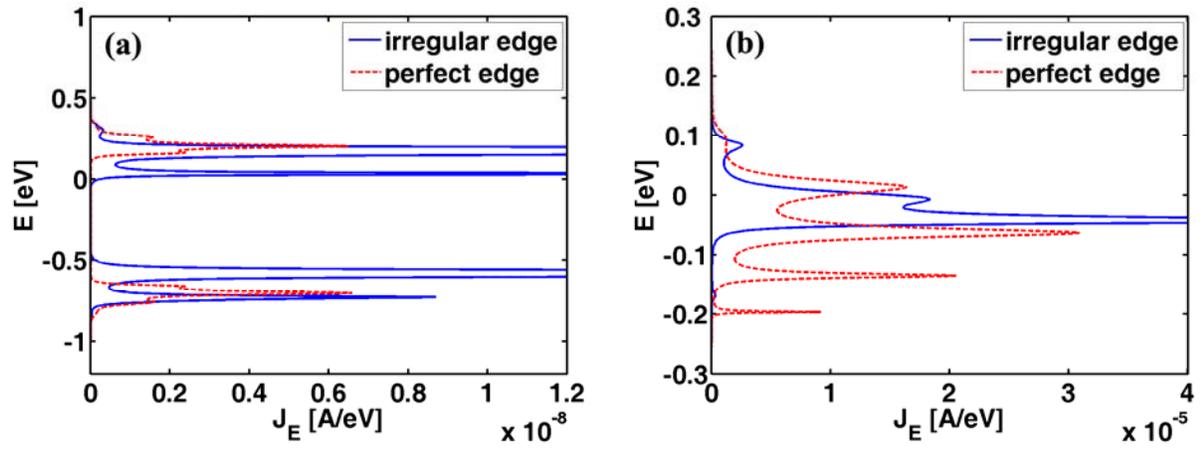

Fig. 4